\documentclass[aps,pra,10pt,twocolumn,showpacs,superscriptaddress]{revtex4-1}

\usepackage{amsmath}
\usepackage{latexsym}
\usepackage{amssymb}
\usepackage{bm}
\usepackage{graphics,epstopdf}
\usepackage{color}
\usepackage{hyperref}

\usepackage{newlfont}
\usepackage{amsfonts}
\usepackage{amsthm}
\usepackage{graphicx}
\usepackage{epsfig}

\usepackage{rotating}
\usepackage{tikz}
													
\usepackage{amsmath}
\usepackage{latexsym}
\usepackage{amssymb}
\usepackage{bm}
\usepackage{hyperref}

\usepackage{newlfont}
\usepackage{amsfonts}
\usepackage{amsthm}
\usepackage{graphicx}
\usepackage{epsfig}
\usepackage{times}

\usepackage{natbib}
\usepackage{caption}
\usepackage{subcaption}

\begin{document}

\title{Quantum Acceleration Limit}

\author{Arun K. Pati}

\affiliation{Center for Quantum Science and Technology,\\
IIIT, Hyderabad, India\\
}

\begin{abstract}
The speed limit provides an upper bound for the dynamical evolution time of a quantum system. Here, we introduce the notion of quantum acceleration limit for unitary time evolution of quantum systems under time-dependent Hamiltonian. We prove that the quantum acceleration is upper bounded by the fluctuation in the derivative of the Hamiltonian. This leads to a universal quantum acceleration limit (QAL) which answers the question: What is the minimum time required for a quantum system to be accelerated from arbitrary initial state to final state?  We illustrate the quantum acceleration limit for a two-level quantum system and show that the bound is indeed tight. This notion can have important applications in adiabatic quantum computing, quantum control and quantum thermodynamics.
\end{abstract}

\maketitle

{\it Introduction.--} In quantum mechanics, the quantum speed limit (QSL) tells us how fast a systems can evolve in time. It was first discovered  by Mandelstam and Tamm (MT) \cite{Mandelstam1945} for quantum systems  undergoing unitary dynamics. At the heart of QSL, the concept of speed of transportation of a quantum system plays a pivotal role. Given an initial state and a final state, the QSL is completely governed by the speed of transportation of the state. The later notion was first introduced by Aharonov and Anandan \cite{Anandan1990} using the Fubini-Study metric. The speed of transportation of the quantum system for arbitrary time evolution was described using the Riemannian metric in Ref. \cite{AKPati03, AKPati04, AKPati05}. The speed of transportation of quantum state is very useful in providing geometric meaning to several quantum phenomenon. The notion of maximal acceleration for a quantum particle in the Euclidean space is bounded by the speed of transportation on the projective Hilbert space \cite{pati01}. The transition probability for a general two-state system was shown to be related to the quantum speed \cite{pati02}. The super current in the Josephson junction was shown to be directly related to the speed of transportation \cite{pati03}.
The QSL described as the maximum evolution speed of the quantum system. It determines the shortest evolution time needed for a quantum system to evolve from an initial state to the target state \cite{Mandelstam1945, Anandan1990,Margolus1998,Levitin2009,Gislason1956,Eberly1973,Bauer1978,Bhattacharyya1983,Leubner1985, Vaidman1992,Uhlmann1992,Uffink1993,Pfeifer1995,Horesh1998,AKPati1999,Soderholm1999,Giovannetti2004,Andrecut2004,Gray2005,Luo2005,Batle2005,Borras2006,Zielinski2006,Zander2007,Andrews2007,Kupferman2008,Yurtsever2010,Fu2010,
Chau2010,Jones2010,Zwierz2012,S.Deffner2013,Fung2013,Poggi2013,Fung2014,Andersson2014,D.Mondal2016,Mondal2016,S.Deffner2017,Campaioli2018}. Over last several years, QSL has been explored for isolated quantum systems. There is an upsurge of interest in its generalisations for open quantum systems \cite{,Mondal2016,Taddei2013,Campo2013,Deffner2013,Pires2016,S.Deffner2020}. Notably, not only we have a lower bound as decided by QSL, but we have an upper bound which is called the reverse quantum speed limit (RQSL) \cite{brij}. The RQSL tells that how slow a quantum system can evolve in time. The reverse quantum speed limit provides us the minimum evolution speed of quantum systems. It sets a bound on the maximum evolution time needed for a quantum state of the quantum system to evolve from a given initial state to the target state. Recently, quantum speed limit has been generalised for arbitrary time-continuous evolution \cite{dimpi}. The notion of speed limit has been generalised for observables in Ref.\cite{brij1}.  The speed limit
concerning the fluctuation of a general observable has been discussed in Ref.\cite{hama}. It has been shown that it is possible to prove the exact quantum speed limit and the MT bound is a special cases of this \cite{brij2}.

In this paper, we ask the question: Does quantum theory put any limit on the acceleration of a quantum system on the projective Hilbert space? Note that this acceleration is not in the ordinary Euclidean space. The speed of transportation of a quantum state is governed by the fluctuation in the Hamiltonian of the system. Even though the unitary time evolution is generated by the Hamiltonian, the speed is not decided by the Hamiltonian alone. Rather the variance in the Hamiltonian decides how fast or how slow the system evolves in time. If the Hamiltonian is time-independent, then the speed of transportation of the state will be time-independent. In this case, we cannot talk about rate of change of speed of transportation. However, if the Hamiltonian is time-dependent, then the quantum acceleration which is nothing but the rate of change of speed of transportation will be non-zero. We will prove that the quantum acceleration is upper bounded by the fluctuation in the rate of Hamiltonian. We illustrate the quantum acceleration limit (QAL) for a two-state system under time-dependent Hamiltonian and show that QAL saturates. Our bound can have several applications in the adiabatic quantum computing, quantum control, quantum battery and quantum thermodynamics.


{\it Bound on quantum acceleration.--} 
Before we prove the main result, we need some background on geometry of quantum evolution. Consider a set of  vectors $\{\psi\}$ of $(D+1)$-dimensional quantum system  that belongs to a Hilbert space $C^{D+1}$. If these vectors are not normalized, we can consider $ \{ \psi/|| \psi ||  \}$ be a set of vectors of norm one belongs to unit normed Hilbert space $\mathcal{L}$.
The state of a quantum system is represented by a ray in the Hilbert
space $\mathcal{H}$. Two normalized state vectors $|\psi\rangle$ and $|\psi'\rangle$ are equivalent if they belongs to same ray in the Hilbert space. Which means  they merely differ by a phase factor (${|\psi'\rangle} \equiv {e^{i\phi}|\psi\rangle}$, where $e^{i\phi}$ $\in$   $U(1)$). The set of rays of $\mathcal{H}$ via a projection map is known as the projective Hilbert space $\mathcal{P}$. The projection map $\Pi:\mathcal{L} \xrightarrow{} \mathcal{P}$ is a principal fibre bundle  $\mathcal{L}(\mathcal{P},U(1),\Pi)$, with structure group $U(1)$. This can be observed by considering the action of the multiplicative group $C^{*}$ of non-zero complex numbers on the space $C^{D+1}-\{0\}$ given by the equivalence relation $(z_{1},z_{2},. ..,z_{D+1})\lambda:= (z_{1}\lambda,z_{2}\lambda,. ..,z_{D+1}\lambda)$ $\forall$  $\lambda$  $\in$ $C^{*}$. This is a free action and the orbit space is the space $\mathcal{CP}^{D}$  of complex lines in the Hilbert space $\mathcal{H}=\mathcal{C}\mathcal{P}^{D+1}$. Thus, we get the principal bundle $C^{*}\xrightarrow{}C^{D+1}-\{0\}\xrightarrow{}\mathcal{CP}^{D}=\mathcal{P}$ in which the projection map associates with each $(D+1)$ tuple $(z_{1},z_{2},. ..,z_{D+1})$ the point in $\mathcal{P}(\cal H) = \mathcal{CP}^{D}$ with the homogeneous coordinates. Thus, any pure quantum state at given instant of time is represented by a point in $\mathcal{P}(\cal H) $ and the evolution of the quantum system is represented by a curve $\Gamma$ in $\mathcal{H}$, which projects to a curve $\hat{\Gamma}$ $=\Pi({\Gamma})$ in $\mathcal{P}$ \cite{AKPati03,AKPati04,AKPati05}.\\

Now, consider the time evolution of a quantum system under time-dependent Hamiltonian $H(t)$. The system evolves according to the Schr\"odinger equation 

\begin{align}
  i \hbar \frac{d}{dt} | \Psi(t) \rangle = H(t) | \Psi(t) \rangle.
\end{align}
As the system evolves in times, it traces a path on the projective Hilbert space. The infinitesimal distance as measured by the Fubini-Study metric is defined as 
\begin{align}
  dS^2 = 4( 1-  |\langle \Psi(t) | \Psi(t+dt) \rangle |^2)  =  \frac{4}{\hbar^2} \Delta H(t)^2,
\end{align}
where $\Delta H(t)^2 = \langle \Psi(t) | H(t)^2 | \Psi(t) \rangle - \langle \Psi(t)| H(t) | \Psi(t) \rangle^2 $ is the fluctuation in the Hamiltonian. The total distance the system travels on the projective Hilbert space $\mathcal{P}(\cal H) $ is then given by
\begin{align}
  S =  \frac{2}{\hbar} \int \Delta H(t) dt.
\end{align}
Thus, the variance in the Hamiltonian has a deep geometric significance as it arises from the geometry of quantum evolution.
The speed of transportation of the quantum system on the projective Hilbert space is defines as \cite{Anandan1990}
\begin{align}
  V_{\cal P}(t)  =\frac{dS}{dt} =  \frac{2}{\hbar} \Delta H(t).
\end{align}
The speed of transportation plays a pivotal role in providing geometric picture to several quantum phenomenon. 

The celebrated MT bound on quantum speed limit again comes from the geometric consideration, i.e., the total distance travelled by the state on the projective Hilbert space is always greater than or equal to the shortest distance connecting the initial and the final states.  If 
$|\Psi(0) \rangle $ is the initial state and $| \Psi(T) \rangle $ is the final state, then
the QSL for a time-independent Hamiltonian, for example, is given by 
\begin{align}
  T\ge  \frac{\hbar S_0}{\Delta H},
\end{align}
where $S_0$ is the shorest distance defines as $\cos S_0/2 = | \langle \Psi(0) | \Psi(T) \rangle |$.

Since $V_{\cal P}(t)$ is the speed of transportation of the state on the projective Hilbert space, we can define the quantum aceleration as 
\begin{align}
  a_{\cal P}(t) =\frac{dV_{\cal P}(t)}{dt} =  \frac{2}{\hbar} \frac{d}{dt} \left( \Delta H(t) \right).
\end{align}
For time-independent Hamiltonian, the quantum acceleration on the projective Hilbert space is identically zero. Therefore, the notion of quantum acceleration will play major role in the time-dependent Hamiltonian systems.

Next, we will show that the uncertainty relation provides an upper bound on the quantum acceleration. To prove this, we can consider square of the quantum speed and take the time derivative. Using the time-dependent Schr\"odinger equation, we can obtain
\begin{align}
  V_{\cal P}(t) \frac{dV_{\cal P}(t)}{dt} =  \frac{4}{\hbar^2} Cov (H, \dot{H}(t) ),
\end{align}
where the covariance of any two observables $A$ and $B$ is defines as 
$Cov (A, B) = \frac{1}{2} (\langle AB + BA \rangle ) - 
\langle A \rangle \langle B \rangle  $ and the averages are taken in the state $| \Psi(t) \rangle$. 
Here, $\dot{H}(t) = \frac{d H(t)}{dt}$. Also, note that for unitary dynamics if $H(t)$ is Hermitian, so also $\dot{H}(t)$.
Now, using 
the Schr\"odinger-Robertson uncertainty relation \cite{rob,sch} for two Hermitian observables $A$ and $B$, we have 
\begin{align}
  \Delta A^2  \Delta B^2 \ge   Cov (A, B)^2 + \frac{1}{4} |\langle[ A, B] \rangle |^2.
\end{align}
Here, if we take $A=H(t)$ and $B= \dot{H}(t)$, we can show that the squared quantum acceleration satisfies the bound
\begin{align}
  a_{\cal P}(t)^2  
  \le \frac{4}{\hbar^2} \Delta \dot{H}(t)^2 - \frac{1}{\hbar^2} 
  \frac{1}{\Delta H(t)^2 } |\langle [H(t), \dot{H}(t) ] \rangle |^2.
\end{align}
If $H(t)$ and $\dot{H}(t)$ commutes or we drop this commutator term, then we have the bound on the quantum acceleration as given by
\begin{align}
  a_{\cal P}(t)^2 \le \frac{4}{\hbar^2} \Delta \dot{H}(t)^2.
\end{align}
This shows that the quantum acceleration for a pure state undergoing unitary dynamics is indeed upper bounded by the fluctuation in the rate of change of the Hamiltonian. Since, by definition the fluctuation of an observable in a quantum state is a positive quantity, we can also express the above bound as 
\begin{align}
  |a_{\cal P}(t)|  \le \frac{2}{\hbar} \Delta \dot{H}(t).
\end{align}
This is one of the central result of our paper. Incidentally, this result was obtained by the present author in 2010, but it was never written up.

Using the above limit, we can prove a universal quantum acceleration limit (QAL) which is independent of the initial and final states. From the above equation, if we integrate on both the sides and define the time average of the fluctuation in the
rate of Hamiltonian and using the inequality $|\int f(t) dt| \le \int |f(t)| dt $, we have 
\begin{align}
 T \ge  \frac{\hbar}{2} \frac{V_{\cal P}(T)}{\Gamma },
\end{align}
where $\Gamma = \frac{1}{T} \int_0^T  \Delta \dot{H}(t) dt$ and $V_{\cal P}(T) = \frac{2}{\hbar} \Delta H(t)|_{t=T}$ is the speed of the quantum system at time $t=T$. We can define $T_{QAL} = \frac{\hbar}{2} \frac{V_{\cal P}(T)}{\Gamma } $ and this provides a minimum time 
required to accelerate the system from arbitrary initial state to arbitrary final state. If we have a time-indepedent Hamiltonian, then $\Gamma =0$ and we can never be able to accelerate a quantum system. This is consistent with QAL as given by Eq.(11). This novel form of quantum acceleration limit will be explored more in future. We expect that the QAL will play a major role in quantum control of time-dependent Hamiltonian systems.

{\it Example and tightness of QAL.--} Here, we illustrate with a simple example for a two-state system such as a spin in time-dependent external field. 
 Consider an atom with two energy levels in time dependent external field. The time-dependent Hamiltonian is given by
\begin{equation*}
    H =  J(t)\sigma_{x},
\end{equation*}
where $J(t)$ is defined as $J(t) = 0 $ for $ t = 0$ and $J(t)$ for $ t > 0$.
Let $|0\rangle$ and $|1\rangle$ are the excited and ground states of the atom.
 If initial state of system is prepared in the state $| \Psi(0) \rangle = |0 \rangle$, then at a later time, we have 
\begin{equation*}
   |\Psi(t)\rangle =\ cos \theta(t) |0\rangle - i \sin \theta(t) |1\rangle,
\end{equation*}
where $\theta(t) = \frac{1}{\hbar} \int_0^t J(t')dt'$. For this two-state system, the fluctuation in the rate of Hamiltonian in the state 
$| Psi(t) \rangle $ is given by $\Delta \dot{H}(t)= \dot{J}(t)$. Therefore, the quantum acceleration limit can be expressed as $a_{\cal P}(t)^2 \le \frac{4}{\hbar^2}\dot{J}(t)^2 $. Thus, in this case, the magnitude of quantum acceleration is upper bounded by the rate of change of the driving field strength. To check the QAL, note that the spin-qubit travels with a speed of transportation $V_{\cal P}(t) = \frac{2}{\hbar} J(t)$ and $\Gamma = \frac{J(T)}{T}$. We can check that for this time-depedent spin Hamiltonian the QAL actually saturates, i.e., equality holds.

 Next, we will show that for any linear parametric coupling time-dependent Hamiltonian, i.e., if $H(t) = \lambda(t) H_0$, where $\lambda(t)$ is the coupling parameter or external controllable parameter and $H_0$ is independent of time, then QAL saturate in any dimensional Hilbert space system. For such a Hamiltonian, the speed of transportation of the system can be expressed as 
 $V_{\cal P}(t) = \frac{2}{\hbar} \lambda(t) \Delta H_0$. Assuming that $\lambda(0) =0$ and $\lambda(t) > 0$ for $t >0$, we can check that $\Gamma = \lambda(T) \frac{ \Delta H_0}{T} $. For this kind of system, indeed QAL saturates, i.e., $T_{QAL} = T$.

 {\it  QAL in adiabatic quantum computing.--} 
 Quantum adiabatic optimization is a promising method for
solving optimization problems using quantum computers. In this approach, one designs a Hamiltonian whose ground state encodes the solution of an optimization problem \cite{farhi,gut}. Then, one can prepare the known ground state of a simple Hamiltonian. Subsequently, we can interpolate the Hamiltonian slowly to reach the final Hamiltonian.
Specifically, consider an initial Hamiltonian, $H_I$, whose ground-state is easy to prepare and 
a final Hamiltonian, $ H_F$ , whose ground-state encodes the solution to the problem at our disposal.
An adiabatic evolution path can be parameterised by $ s(t) $ where $s(0) = 1$ and $s(T ) = 0$ with $T$ being the total run time. The time-dependent Hamiltonian
can be expressed as 

\begin{align}
H(s(t)) = H(t) = s(t)H_I + (1 - s(t))H_F, 
\end{align}
where $s(t) = (1- \frac{t}{T})$ with $H(0) = H_I$  and $H(T) = H_F$. The quantum system evolves according to the Schr\"odinger equation with time-dependent solution parametrised by $s(t)$, i.e., $|\Psi(s(t)) \rangle$. 
The quantum adiabatic theorem tells us that the final state of the quantum system will be close to the ground-state of $H_F$ which encodes the solution to
the problem.

Let $| Psi(0) \rangle  = | E_0(0) \rangle $, where $| E_0(0) \rangle $ is the ground state of $H(0)$. Adiabatic theorem tells us that for large $T$, we can have $|\langle E_0(1) | \Psi(T) \rangle  \rightarrow 1$. 
The total run time $T$ depends on the spectrum of the Hamiltonian. In the past, there are results proving the lower bound on the run time. Here, we will prove a new lower bound that involves both the quantum speed and quantum acceleration. On using Eqs. (7) and (9) and noting that 
$[H(s(t)), \dot{H}(s(t)) ] = \frac{1}{T}[H_I, H_F]$, we have 

\begin{align}
    T \ge max_{s(t) \in [0,1]} \frac{1}{2 \Delta H(s(t)) } 
    \frac{ |\langle [H_I, H_F] \rangle |}{ \sqrt{  \Delta \dot{H}(s(t))^2 - \hbar^2 a(s(t))^2 /4 } }.
\end{align}

Additionally, we can prove an upper bound for the run time $T$ using the bound on the quantum acceleration. Note that the fluctuation in the rate of Hamiltonian is given by 
\begin{align}
 \Delta \dot{H}(t) =\frac{1}{T} \Delta(H_F - H_I).
\end{align}
Now, using the sum uncertainty relation \cite{patisahu,mac} for any two observables, we have $\Delta(H_F - H_I) \le \Delta H_F + \Delta H_I$. Then, using Eq.(10), we have an upper bound for the total run time as given by 

\begin{align}
   T \le \frac{2}{\hbar } \frac{1}{ V_{\cal P}(T) } \int_0^T ( \Delta H_I(t)  + \Delta H_F(t) )~ dt.
\end{align}
Thus, it is possible to lower the total run time by reducing the fluctuation in the initial and final Hamiltonians. Also, if we can increase the speed of adiabatic quantum system during the interpolation then it is possible to reduce the total run time.
The above bound is consistent with the intuition that if we drive the system by counter adiabatic Hamiltonian, i.e., follow the method of short-cut to adiabaticity, we can indeed reduce the run time \cite{del,muga}. We will explore in more detail the quantum acceleration limit in the context of short-cut to adiabaticity.

{\it Conclusion.--} The notion of quantum speed limit (QSL) has played a significant role in quantum mechanics, quantum computing, quantum control and variety of other areas over last several decades. Here, we have shown that the fundamental uncertainty relation provides an upper bound on the acceleration of a quantum state on the projective Hilbert space. The quantum acceleration limit is completely governed by the fluctuation in the rate of time-dependent Hamiltonian.  We have shown that for any linear coupling externally controllable time-depedent parametric Hamiltonian, QAL actually saturates. This demonstrates that QAL is indeed tight for two-level systems and a class of higher dimensional systems. We have proved new lower and upper bounds on the run time of adiabatic quantum computing. Our result can have deep implication in adiabatic quantum computing, quantum control and quantum thermodynamics where time-dependent Hamiltonian is used to drive the system. In future, it will be interesting to prove the quantum acceleration limit for open system dynamics.

\vskip 2cm

{\it Note:} After completion of our work, we noticed that quantum acceleration on the projective Hilbert space is also explored
in a recent preprint by Paul M. Alsing and Carlo Cafaro, arxiv:2311.18470.

\end{document}